\documentclass[aps,prl,longbibliography,twocolumn,superscriptaddress,showpacs]{revtex4-1}

\usepackage{CJK}
\usepackage{graphicx}
\usepackage{amsmath}
\usepackage{dcolumn}
\usepackage{bm}
\usepackage[normalem]{ulem}
\usepackage{xcolor}
\usepackage{hyperref}

\usepackage{enumitem,kantlipsum}
 
\usepackage{epstopdf}

\newcommand{\rf}{\rho_{ref}}
\newcommand{\ms}{M_{\rm spinodal}}
\newcommand{\mb}{M_{\rm binodal}}

\newcommand{\rc}{\rho_{c}}

 \newcommand{\nn}{\nonumber}

\newcommand{\vb}{\langle\bv\rangle}

\newcommand{\bew}{\begin{widetext}}
\newcommand{\ew}{\end{widetext}}

\newcommand{\bq}{\mathbf{q}}
\newcommand{\bv}{\mathbf{v}}

\newcommand{\br}{\mathbf{r}}

\newcommand{\bvp}{\mathbf{v}_\perp}

\newcommand{\hx}{\hat{x}}
\newcommand{\hp}{\hat{p}}

\newcommand{\sep}{ \ \ \ , \ \ \ }

\newcommand{\beq}{\begin{equation}}
\newcommand{\eeq}{\end{equation}}
\newcommand{\beqn}{\begin{eqnarray}}
\newcommand{\eeqn}{\end{eqnarray}}
\newcommand{\pp}{\partial}

\begin{document}

\title{ Following your nose: Autochemotaxis and other mechanisms for spinodal decomposition in flocks}
\author{Maxx Miller}
\email{maxxm@uoregon.edu}
\affiliation{Department of Physics and Institute for Fundamental
	 Science, University of Oregon, Eugene, OR $97403$}
\author{John Toner}
\email{jjt@uoregon.edu}
\affiliation{Department of Physics and Institute for Fundamental
	 Science, University of Oregon, Eugene, OR $97403$}
 
\begin{abstract}
We develop the hydrodynamic theory of dry, polar ordered, active matter (``flocking") with autochemotaxis; i.e.,
self-propelled entities moving in the same direction, each emitting a substance which attracts the others (e.g., ants). We find that sufficiently strong autochemotaxis leads to an instability to phase separation into one high and one low density band. This is very analogous to both equilibrium phase separation, and ``motility induced phase separation" (``MIPS") and can occur in flocks due to {\it any} microscopic mechanism (e.g., { sufficiently strong} attractive interactions) that makes the entities cohere.
\end{abstract} 
\maketitle

Two of the principle mechanisms of {\it ordered} biological motion are ``taxis"\cite{KS1, KS2, KS3, KSReview, KSCollective} and ``flocking"\cite{Vicsek,TT1, TT2, TT3, TT4, rean} . ``Taxis" is motion directed preferentially along the gradient of some quantity.  In ``autochemotaxis", this quantity is the concentration of a substance secreted by the moving creatures themselves. This can lead to aggregation of those creatures\cite{KS1, KS2, KS3, KSReview, KSCollective}.

``Flocking" is the collective coherent motion of a large number of organisms (e.g., flocks of birds) that occurs when creatures follow their neighbors. Strikingly, even purely short ranged interactions generate long-ranged order in an arbitrarily large flock in two dimensions\cite{TT1,TT2,TT3,TT4, rean}, which is forbidden in equilibrium systems by the Mermin-Wagner-Hohenberg Theorem\cite{MW}.
 
Hydrodynamic theories of autochemotaxis\cite{KS1, KS2, KS3, KSReview, KSCollective} and flocking\cite{TT1, TT2, TT3,TT4, rean} {\it separately} already exist. In this paper, we develop and study a hydrodynamic theory for systems in which {\it both} effects are present. Specifically, we consider active, self-propelled particles (hereafter called ``boids") moving over a frictional substrate, so that momentum is not conserved and time reversal symmetry is broken. The underlying dynamical rules are assumed to be rotation and translation invariant. This combination, as we show in this paper, can lead to trails (e.g., ant trails).

More specifically, we investigate autochemotaxis in an ``active homogeneous polar ordered state", which spontaneously breaks the underlying rotation invariance by having all of the boids move, on average, in the same direction (i.e., ``flock"). The homogeneity of the state means that translation invariance is {\it not} broken.

\begin{figure}
    \centering
   \includegraphics[width=0.9\linewidth]{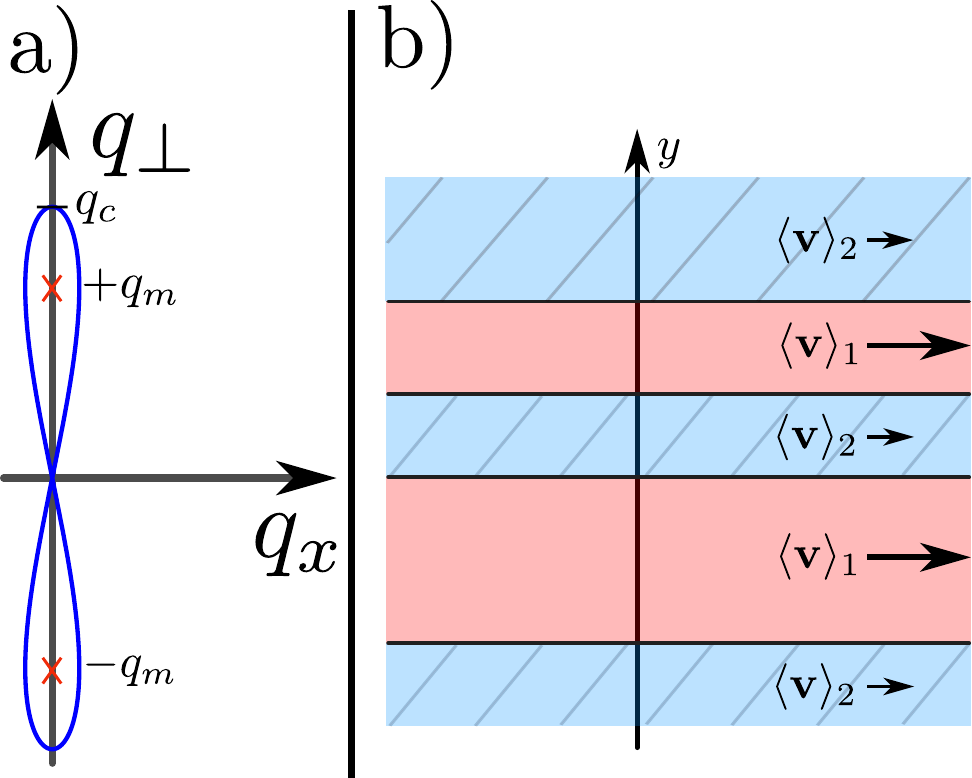}
    \caption{ a) The region of instability in $\bq$ space, for autochemotaxic systems very close to the instability threshold on the unstable side. The instability  is towards forming ``bands" in real space running {\it parallel} to the direction of mean flock motion, as illustrated in b) and figure \ref{Band2}. b) The ``band" structure of the instability at intermediate times. The density is only modulated along one of the directions (which we call $y$, and which is indicated in the figure) perpendicular to the direction $\hx$ of mean flock motion. The mean speeds within each band also differ.}
    \label{q-unstable}
\end{figure}

\begin{figure}
    \centering
 \includegraphics[width=\linewidth]{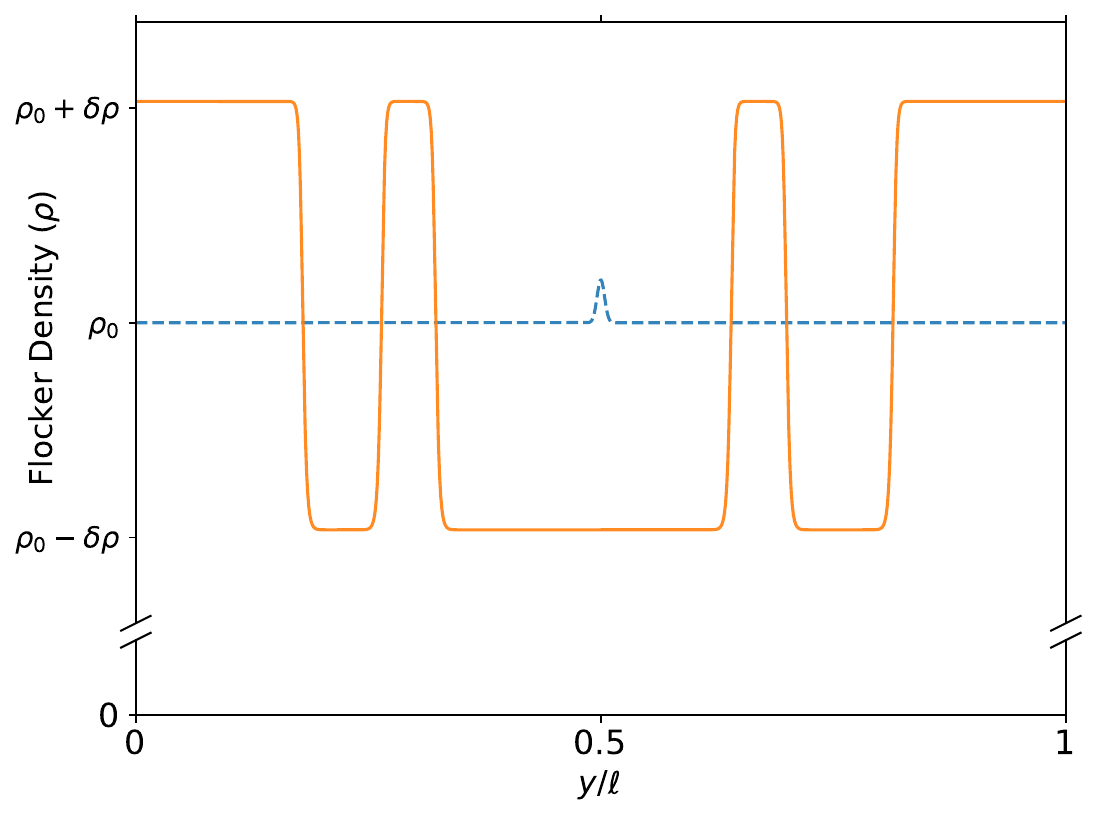}
    \caption{A plot of  the initial (dashed blue curve) and final (solid orange curve) density for a typical numerical solution of the hydrodynamic equations for 1d configurations, as described in the text, at intermediate times. 
        Here $\ell$ is the linear spatial extent of  our system in the $y$-direction (which is periodic).}
    \label{Band2}
\end{figure}

A linear stability  analysis of the hydrodynamic theory  shows  that, like disordered active systems undergoing only autochemotaxis \cite{KSCollective}, the flock  becomes unstable when the autochemotaxis is sufficiently strong. The instability, near threshold, is extremely anisotropic. Specifically, plane wave modulations of the density with wavevector $\bq$ are only unstable in  a narrow region of region of $\bq$ space, extended in the direction perpendicular to the direction of flock motion, as illustrated in figure \ref{q-unstable}a. As a result, the instability is towards forming ``bands" in real space running {\it parallel} to the direction of mean flock motion (denoted by $\hx$), as illustrated in figures \ref{q-unstable}b and \ref{Band2}. 

{  The instability we find here is completely different from the ``banding instability" \cite{BLee, Banding1, Banding2, rean}, which occurs in a completely different region of parameter space is driven by completely different physics\cite{BLee, Banding1, Banding2, rean}, and is characterized by the formation of bands {\it perpendicular} to the direction of mean flock motion.}

{ The results we obtain herein also apply to {\it any} dry polar ordered active matter system in which {\it any} sufficiently strong microscopic mechanism drives the inverse compressibility negative, since any such system will be described, at the longest length and time scales, by a Toner-Tu model with a negative inverse compressibility, which is the basis for all of our predictions here.
}

Motivated by the geometry of the instability, we  analytically and numerically solved our equations of motion for ``one-dimensional" configurations, by which we mean configurations in which all of the fields depend { {\it on only one of the $d-1$ Cartesian coordinates} } perpendicular to $\hx$ ($d$ being the dimension of space).  

Our numerical solutions show that these bands quickly evolve into a set of well-separated parallel ``plateaus''. In each plateau, the density is almost constant, as illustrated in figure \ref{Band2}. The evolution then continues via  band merging, ultimately leading to full phase separation into a single high density band, and a single low density band. 

\begin{figure}
        \centering
        \includegraphics[width=0.8\linewidth]{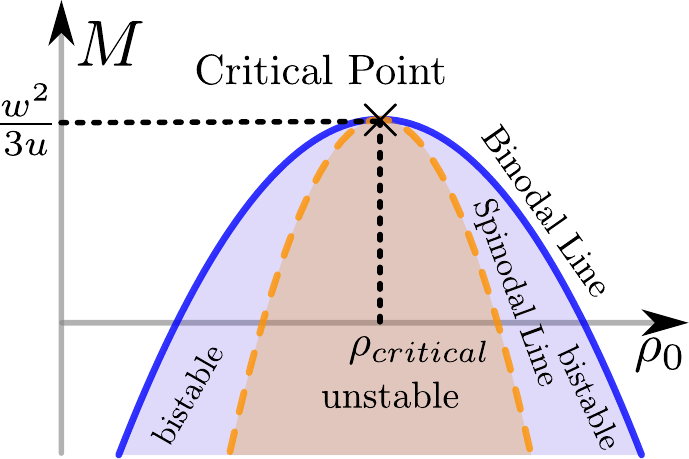}
        \caption{Phase diagram of flock phase-separation. Solid blue and dashed orange lines are the binodal and spinodal lines respectively. The orange filled region, labeled ``unstable", corresponds to the coexistence of the 'liquid' (high density) and 'gas' (low density) phases.}
        \label{spinodal}
\end{figure}

Our analytic steady-state solution  implies the phase diagram shown in figure (\ref{spinodal}), in which the vertical axis is a model parameter defined later which decreases with increasing autochemotaxis. This analytic solution is asymptotically valid sufficiently close to 
the critical point in  this figure. In the region above the curve labelled ``Binodal line", the uniform state is the only stable steady state of the flock. Below the curve labelled ``Spinodal line", only the phase separated state is stable. Between these two curves, {\it either} state can be stable.  

There is a fundamental dissimilarity between our result and equilibrium liquid-vapor phase separation. In equilibrium, a homogeneous state in this part of the phase diagram would be {\it meta}-stable; that is, it would occupy a {\it local}, but not a {\it global} minimum in the energy landscape. If fluctuations were sufficiently large, it could reach the global minimum, and phase separate. Because flocks are {\it non-equilibrium} systems, we have no analog to this energy argument, and thus can only say that the homogeneous and phase separated states are locally stable between the binodal and spinodal curves. There is no criterion known to us that unambiguously identifies  one state as truly stable, and the other as meta-stable. We therefore simply call this region the bistable region. 

Despite this distinction, those familiar with equilibrium phase separation will see the similarities with our system, although ours also differs from the equilibrium case both because of its strong anisotropy, and because of non-equilibrium effects.

One of these effects is to change  the familiar ``common tangent construction" for  the two coexisting densities to an {\it  uncommon} tangent construction. This is very similar to the behavior found in  motility-induced phase separation (MIPS) in {\it disordered} active systems \cite{mipsreview}. A common tangent construction {\it does} become asymptotically exact as  the system  approaches the  critical point.

Our treatment of this phase separation is entirely ``mean-field": that is, it ignores fluctuations in the local density and velocity. Fluctuations in the density are well-known to radically change the scaling behavior of the density and dynamics near the critical point in equilibrium systems\cite{chaikin, Ma}. Furthermore, even  {\it stable} (i.e., non-phase-separating) flocks experience  divergent fluctuation corrections to their dynamics\cite{TT1,TT2,TT3,TT4, rean}. We therefore expect the true scaling behavior close to the critical point we find here to differ from our mean
field results.

The hydrodynamic theory combines the Keller-Segel  equations\cite{KS1, KS2, KS3, KSReview, KSCollective} for autochemotaxis with the Toner-Tu equations\cite{TT1, TT2, TT3, TT4, rean} into a unified set of hydrodynamic equations for autochemotaxic flocking. This a continuum model which takes as its variables three fields: the number density of boids $\rho(\br,t)$,   the boid velocity field $\bv(\br,t)$, and the concentration $\eta(\br,t)$ of the chemical ``signal" (or ``chemo-attractant"). The hydrodynamic equations governing the time evolution of these fields are constrained by the symmetries (rotation and translation invariance) and  conservation law (boid number) of  the underlying dynamics to take the form \cite{vdef}:  

\bew
\beqn
\partial_t\bv + \lambda_1(\bv\cdot\nabla)\bv + \lambda_2(\nabla\cdot\bv)\bv + \lambda_3\nabla(|\bv|^2) &=& 
    U(|\bv|,\rho, \eta)\bv - \nabla P_1(|\bv|,\rho, \eta) - \bv(\bv\cdot\nabla P_2(|\bv|,\rho, \eta))
     \nn\\
     &+& D_B\nabla(\nabla\cdot\bv) + D_T\nabla^2\bv + D_2(\bv\cdot\nabla)^2 \bv , \label{KSTTvIntro}
     \eeqn
     \beqn
         \pp_t\rho&=& \nabla \cdot\bigg( k_1(|\bv|, \rho, \eta) \nabla \rho - k_2(|\bv|, \rho, \eta)\rho\nabla \eta +k_{1a}(|\bv|, \rho, \eta)\bv (\bv\cdot\nabla \rho) - k_{2a}(|\bv|, \rho, \eta)\rho\bv (\bv\cdot\nabla \eta)\bigg)-\nabla\cdot(\bv\rho),
    \label{KSTTrhoIntro}
    \eeqn
    \beqn
    \pp_t\eta&=& D_\eta\nabla^2\eta + k_6(|\bv|,\rho,\eta)
    -\lambda_{\eta1}(\bv\cdot\nabla)\eta - \lambda_{\eta2}(\nabla\cdot\bv)\eta { -\zeta_\eta\nabla\cdot(\bv \rho)}+D_{\eta\rho} \nabla^2\rho+D_{\eta v}\nabla^2|\bv|\nn\\
    &+&    D_{\eta a} \nabla\cdot[\bv(\bv\cdot\nabla)\eta]+D_{\eta\rho a} \nabla\cdot[\bv(\bv\cdot\nabla)\rho]+D_{\eta v a} \nabla\cdot[\bv(\bv\cdot\nabla)|\bv|]\,.\label{KSTTetaIntro}
     \eeqn
     \ew
Here, $k_1$ is the motility of the boids, $k_2$  the sensitivity to the chemical signal ,  and  $k_6$  describes the production and decay of the chemical signal \cite{HillenPainterCompsci}.  We have  introduced anisotropic analogs $k_{1a}$ and $k_{2a}$ of $k_1$ and $k_2$, since we expect the response {\it along} the direction of the flock velocity $\bv$ to be different, in general, from the response perpendicular to that direction. The numbering of the $k$'s follows the convention of Keller and Segel\cite{KS1, KS2, KS3, KSReview, KSCollective, HillenPainterCompsci}.

Our flock is on a frictional substrate, which destroys  Galilean invariance and momentum conservation, and allows  the friction-like $U(|\bv|,\rho,\eta)$ term. Because our system is active and has a moving steady state, $U(|\bv|,\rho,\eta)$ should be positive for small $|\bv|$, and negative for large $|\bv|$, thereby ensuring that the flock settles down to a steady state with a non-zero average velocity $\langle\bv\rangle$.

For a discussion of the physical significance of the other terms in 
(\ref{KSTTvIntro}) and (\ref{KSTTrhoIntro}), see\cite{TT1, TT2, TT3, TT4, rean}.

The quantities  $k_{1,2,1a,2a,6}$, $P_{1,2}(|\bv|,\rho,\eta)$, $U(|\bv|,\rho,\eta)$, $D_{B,T,2}$, and $\lambda_{1,2,3}$ are in general functions $|\bv|$, $\eta$, and $\rho$. They can {\it not} depend of the {\it direction} of $\bv$ due to rotation invariance. We expand all of these in powers of the departures of $|\bv|$, $\eta$,  and  $ \rho$ from their  values in an assumed  uniform steady state     $\rho(\br,t) = \rho_0$, $ \eta(\br,t)= \eta_0$, and $ \bv = v_0\hat{x}$. Here the direction $\hx$ is spontaneously selected by the system, thereby spontaneously breaking the  rotational symmetry of the underlying dynamics.

We can simplify these equations by ``eliminating"  the fields $\delta v_x$ and $\delta\eta$, which  are ``fast" in the sense that their time derivatives do not vanish in the limit of extremely slowly spatially varying fields. See \cite{ALP} for details. This ``elimination'' gives a closed pair of equations for $\rho$ and the projection $\bvp$ of the velocity field perpendicular to the direction $\hx$ of mean flock motion. These equations take {\it exactly} the form of the Toner-Tu equations\cite{rean} or  flocks {\it without} chemotaxis, which are identical in {\it form} to equations  (\ref{KSTTvIntro}) and (\ref{KSTTrhoIntro}) with the chemical signal concentration $\eta$ set to zero,  with one crucial difference:  the effective inverse compressibility (or, equivalently, the bulk modulus) $B\equiv{\pp P_1\over\pp\rho}\bigg|_{\rho_0}$ in these equations is reduced by the autochemotaxis, and can be driven {\it negative} for sufficiently strong chemotaxis. The development of a negative compressibility is also the mechanism which drives equilibrium liquid-gas phase separation; however, the nature of the ensuing instability here is different, as we'll see.

To investigate this instability, we linearized the equations of motion, and determined the complex eigenfrequencies of Fourier modes with wavevector $\bq$. We found that these modes become unstable when $B$ become negative, for $\bq$'s inside the $\bq$ space lemniscate depicted in figure \ref{q-unstable}a, whose shape is given by
\begin{align}
     q_x^2 = \frac{|B|q_\perp^2-D_{\rho\perp}D_{_{L\perp}}q_\perp^4}
      {V^2},
     \label{inst bound alt}
 \end{align}
where  $V$ and $D_{\rho\perp,L}$ are constant model parameters with the dimensions of speed and diffusion constants, respectively. (All of these parameters can be expressed in terms of those of the original equations of motion (\ref{KSTTvIntro}-\ref{KSTTetaIntro}); the detailed expressions for them are given in the ALP\cite{ALP}). Here $q_x$ and $q_\perp$ are the magnitudes of the projection of the wavevector $\bq$ along, and perpendicular to, the direction $\hx$ of mean flock motion, respectively.

Note that as $B\to0^-$ from below, this lemniscate becomes vanishing small, and infinitely narrow. Specifically, its length along the $\perp$ direction vanishes like $\sqrt{|B|}$, while its width along $q_x$ vanishes like $|B|$, as the  reader can  verify for herself from (\ref{inst bound alt}). Another measure of the narrowness of the lemniscate is the ``opening angle" $\theta_c$ of the self-crossing at  $\bq={\bf 0}$, which vanishes like $\sqrt{|B|}$.

The most unstable mode lies at $q_x=0$; i.e., wavevector {\it exactly perpendicular} to the direction of mean flock motion $\hx$. 

The origin of this instability, and the reason it is confined to modes nearly orthogonal to the direction of mean flock motion,  is the competition between convection and pressure forces.
 
By ``convection", we mean that, in the absence of pressure forces and diffusion,  any instantaneous local configuration of $\rho(\br, t)$ or $\bv(\br,t)$ is simply be transported along the direction of flock motion at speeds   $v_\rho$ and $v_v$ respectively. (These speeds can be derived from equations (\ref{KSTTvIntro}- (\ref{KSTTetaIntro}), as is done explicitly in the ALP\cite{ALP}.) The two speeds, $v_\rho$ and $v_v$, are not, in general, equal to one another, nor to the mean speed $|\vb|$ of the flock, due to the aforementioned absence of Galilean invariance. Indeed, their {\it inequality} plays a crucial role in limiting the instability.

To understand this, consider the evolution of a small fluctuation in which both the  density and velocity depart from their mean values in a band perpendicular to the direction of mean flock motion, as illustrated in figure \ref{allbands}. This corresponds to a fluctuation with wavevectors making an angle $\theta_\bq=0$ with the mean velocity $\vb$. The fluctuation in the density creates a pressure gradient, which, because the inverse compressibility $B$ is negative, causes the velocity field to increase in the direction that tends to {\it increase} the fluctuation in the density. That is, if the density is {\it higher} in our band, the pressure forces tend to increase the velocity {\it into} the band, making the density grow even more. The opposite will happen if the density is lower than average in our band: pressure forces will tend to make the velocity flow {\it out}, carrying boids out of the band, {thereby reducing the density even further.
 
This positive feedback mechanism is responsible for the familiar instability of negative compressibility equilibrium systems. In {\it our} system, however, this mechanism must compete against the aforementioned convection. For the band direction just discussed, convection stabilizes the system by separating the band of perturbed velocity from the band of perturbed density, because they propagate at different speeds. This is illustrated in the figure \ref{allbands}a. Once the velocity fluctuation is separated from the density fluctuation, both simply decay away by diffusion. Therefore, this orientation of the bands is stable.

Now consider the other extreme case, in which the band runs {\it parallel} to the direction of mean flock motion, as illustrated in figure \ref{allbands}b. Now the relative displacement of the velocity and density fluctuations makes no difference: the two bands of each still overlap, since the bands are infinitely extended in the direction of that relative motion. Hence, the velocity fluctuation continues to sit right on top of the density fluctuation, allowing the pressure induced instability to occur.

We can even derive the opening instability angle $\theta_c$ by this argument, as we show in the ALP\cite{ALP}. 

\begin{figure}
    \centering
   \includegraphics[width=\linewidth]{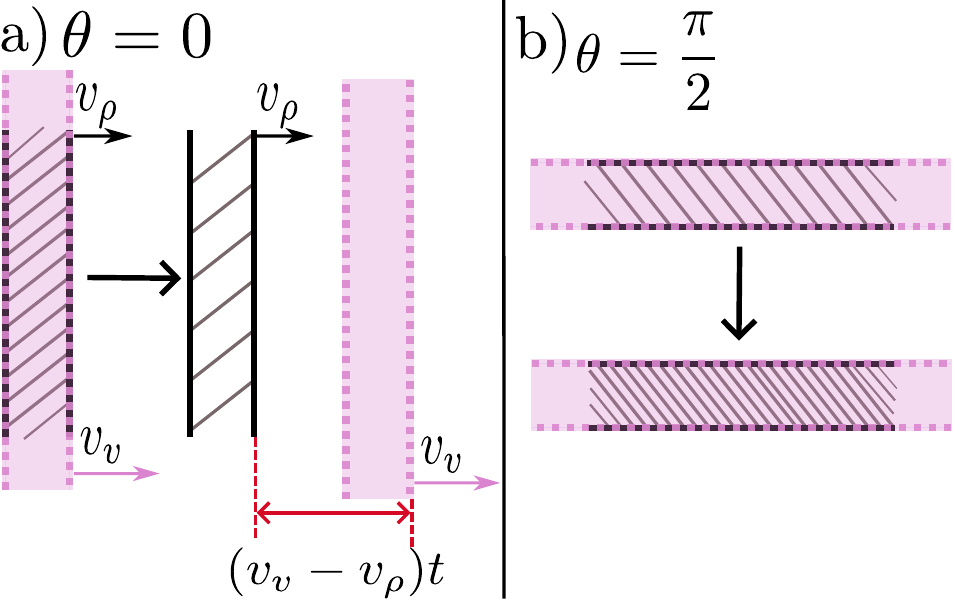}
    \caption{Heuristic illustration of why the bands always form perpendicular to the direction of mean flock motion. See the text after equation  (\ref{inst bound alt}) for explanation.
      }
    \label{allbands}
\end{figure}

We'd now like to determine the final steady state of the system when the instability occurs. Since the instability is sharply focused in the directions perpendicular to the bulk motion, we will seek solutions that depend only on one Cartesian component of position, with that component perpendicular to $\hx$, the direction of mean flock motion. We'll call this direction $y$. Specifically, we'll search for solutions of the form 
\beq
\bvp = v_y(y,t) \hat y \sep
\rho = \rho_{ref} + \Delta\rho(y,t) \,,
\label{1d}
\eeq
 where we expand around a ``reference density" $\rf\ne$ the mean density $\rho_0$ , to be able to treat the behavior of flocks with different $\rho_0$'s in a single unified framework.

Since our system is linearly unstable, we must go beyond linear order in our expansion to find a stable state. It proves to be sufficient to expand our equations of motion  to {\it quadratic} order in the fluctuations $v_y(y,t)$,  and {\it cubic} order in  $\Delta\rho(y,t)$. The most important non-linearities arise from the expansion of $P_1(\rho)$ to third order in  $\Delta\rho$: 
 \beq
 P_1 = P_0+M\Delta\rho + w \Delta\rho^2 + u \Delta\rho^3
 \label{Pexp}
 \eeq
 
We'll begin by determining the {\it stability boundary} in the $\rho$-$M$ plane. As discussed above, and shown in the ALP\cite{ALP}, this boundary is where the inverse compressibility vanishes; that is, when
\beq
B\equiv\left({dP_1\over d\rho}\right)\bigg|_{\rho_0}=0 \,.
\label{stab cond P}
\eeq
We will define the value of $M$ that satisfies this condition to be $M_{spinodal}$, since it is the precise analog of the spinodal line in equilibrium phase separation\cite{chaikin}: that is, the line on which the uniform density state becomes unstable.

Inserting our expansion (\ref{Pexp}) into the condition (\ref{stab cond P}) gives the spinodal line:
\beq
\ms=M_c-3u(\rho_0-\rho_c)^2 \,,
\label{ms}
\eeq
where we've defined the ``critical" value $M_c$ of $M$ and the ``critical" density as
\beq
M_c\equiv{w^2\over3u} \sep
\rc\equiv\rf-{w\over3u} \,.
\label{CP}
\eeq
The critical values $M_c$ and $\rc$ of $M$ and $\rho$ prove to be the analogs of the critical temperature and critical density of an equilibrium phase separating system.

To determine the final steady state of the flock, we looked for steady state solutions to our equations of motion that have two plateaus of constant density, separated by a localized interface. As explained in detail in the ALP\cite{ALP}, we found that such solutions with mean density $\rho_0$  exist only if $M$ is below $M_{binodal}$, where
\beq
\mb=M_c-u(\rho_0-\rho_c)^2 \,.
\label{ms}
\eeq
Note that this is {\it above} the ``spinodal" line (\ref{ms}). Hence, between the two lies a region in which both the uniform state, and the phase separated state, are stable.

We also found that sufficiently close to the critical point (\ref{CP}), the two  plateau densities can be obtained from a ``common tangent construction" on the effective ``Free energy"  $F$ defined by $P_1\equiv{dF\over d\rho}$, which implies
\beq
F(\rho)=F_0+P_0\Delta\rho+{M\over2}\Delta\rho^2 +{w\over3} \Delta\rho^3 + {u\over4} \Delta\rho^4 \,.
\label{fsol}
\eeq
That is, the two plateau densities $\rho_\pm$ are the densities at which a plot of $F(\rho)$ has a common tangent. This is illustrated in figure { (\ref{TCDemo}a)}.

This analogy to equilibrium statistical mechanics {\it only} applies close to the critical point (\ref{CP}). As also explained in detail in the ALP\cite{ALP}, we  found that, as we move away from the critical point, the densities are determined by an ``{\it un}common tangent construction": the slope of the free energy (\ref{fsol}) remains the same at both coexisting densities, but the tangents to the free energy curve at those two densities do {\it not} coincide, as illustrated in figure { (\ref{TCDemo}b)}. In the ALP\cite{ALP}, we  calculate the offset of these two tangents, thereby determining the two coexisting densities.

\begin{figure}
        \centering
        \includegraphics[width=0.8\linewidth]{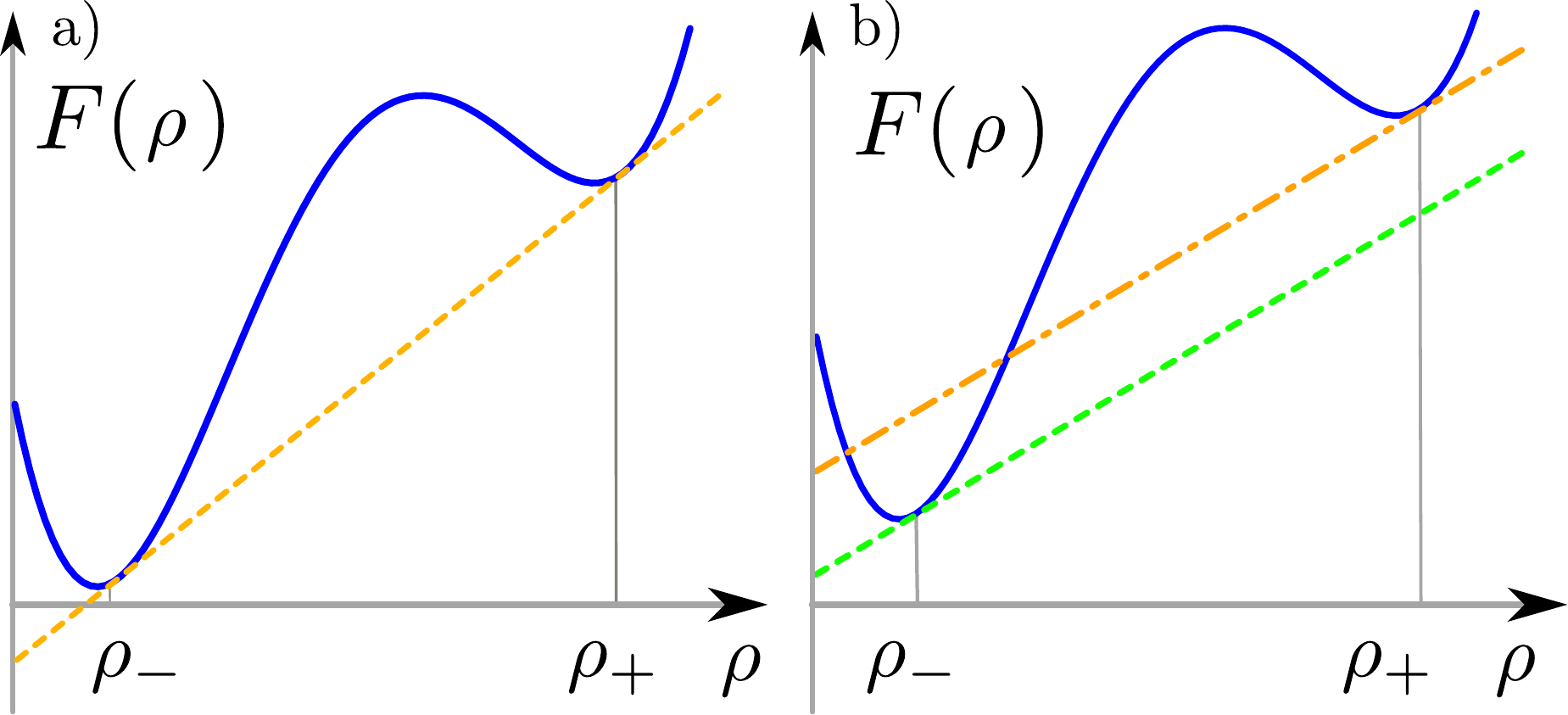}
        \caption{{ a) Illustration of the common tangent construction. b) Illustration of the {\it un}common tangent construction.} }
        \label{TCDemo}
\end{figure}

Finally, we note that since the mean local speed of the flock should depend on the local density, which is different in the two plateaus, the speeds in each plateau, in general, will be different as well, as shown in the second panel of figure (\ref{q-unstable}).

All of these predictions are fully borne out by numerical solution of the time-dependent equations of motion in a periodic space,  as illustrated in figure \ref{Band2}, and in the animation Movie (1) in the supplemental materials. 

All of our simulations started with initial conditions in which the departure from a uniform state   was small. Furthermore, we chose  the mean density $\rho_0$ to be the critical density $\rc$, so our simulations were the analog of a ``critical quench" right through the critical point (\ref{CP}).
  
At small times this system quickly forms into multiple well-defined plateaus, with density and velocity profiles that agree well with our analytic solution. Adjacent plateaus merge quickly, while distant plateaus do not merge on the time scales of our simulations.
 
This slowing down is a consequence of the exponential approach to the uniform plateau state of both the density and the velocity. It is only the overlap of these exponential tails that leads to any interaction at all between neighboring interfaces. As a result, the attraction between neighboring interfaces falls off exponentially with their separation. Hence, the time for neighboring domains of high density to merge also grows exponentially with their separation.

We believe that this slowing down is an artifact of both the one-dimensionality and the noiselessness of our simulations. We expect that in a noisy system, the bands will begin to undulate, causing them  
to bump into their neighbors, at which point the bands can start to merge. We expect this mechanism to reduce the phase separation time to 
grow only algebraically with system size.

In conclusion, we have shown that a dry active autochemotaxic flock, or, more generally, {\it any} flock with { sufficiently strong} attractive interactions { (i.e., strong enough to} drive the inverse compressibility negative)}between the flockers, can become unstable to the formation of density bands running parallel to the mean direction of motion of the flock. This behavior may be connected to the formation of, e.g., unidirectional ant trails. 
The final state of this anisotropic instability is a single high density plateau and a single low density plateau. The coexisting plateau densities of this non-equilibrium phase separation, near its ``critical point'', are determined by a common tangent construction on an effective free energy, while further from the critical point, an {\it un}common tangent construction must be used, as in, e.g, ``motility-induced phase separation" (``MIPS")\cite{mipsreview}. Our analytical theory is in agreement with long-time evolution numerical solutions, even far from the critical point.
    
Our work has focused on noiseless dry active autochemotaxic flocks. Two possible extensions are first, including fluctuation effects on phase separation,  and  second, developing a theory of flock interface dynamics to  understand the late stage phase separation dynamics.

\section{Acknowledgements}
We thank Tristan Ursell and Justin Kittel for stimulating discussions of their experiments on the dynamics of ant trails; those experiments and  discussions motivated our work. We also thank the MPIPKS for their hospitality and financial support while a portion of this work was underway. Finally, we thank Ants for their interesting behavior.

\bibliographystyle{apsrev4-1}

\begin{thebibliography}{21}
\bibitem [{vde()}]{vdef}%
  \BibitemOpen
  \href@noop {} {}\bibinfo {note} {Note that, in contrast to the approach of
  the Toner-Tu equation, the field $v$ we are using in equation (2) is not the
  ``true'' or full velocity field $\bv_{\rm true}$, in the sense that it is not
  the field that is related to the number current of birds by the exact
  continuity equation $ {\bf J}=\rho\bv_{\rm true}$. Instead, we have chosen to
  work with a velocity field defined by $\bv=v_0{\bf \hp}$, where ${\bf \hp}$
  is the local ``polarization'''; that is, a unit vector along the direction in
  which the local flockers are pointing. This fact then requires that we
  include the symmetry allowed k1, k1a, k2 and k2a terms. See the Associated
  Long Paper \cite{ALP} for a further discussion.}\BibitemShut {Stop}%
\end{thebibliography}

\end{document}